\documentclass[reprint,superscriptaddress,floatfix,prl]{revtex4-1}
\usepackage{graphicx}
\usepackage{textcomp}
\usepackage{units}
\usepackage{siunitx}
\usepackage[normalem]{ulem}

\usepackage{amsmath}
\usepackage{amsfonts}
\usepackage{amssymb}
\usepackage[utf8]{inputenc}

\usepackage[usenames,dvipsnames,svgnames,table]{xcolor}
\usepackage[normalem]{ulem}

\newcommand{\Tc}{T_\mathrm{c}}

\begin{document}


\title{Microscopic charging and in-gap states in superconducting granular aluminum}


\author{Fang Yang}
\email{francoyang1988@163.com}
\affiliation{Institute~for~Nanoelectronic~Devices~and~Quantum~Computing,~Fudan~University,~200438~Shanghai,~China}
\affiliation{Physikalisches~Institut,~Karlsruhe~Institute~of~Technology,~76131 Karlsruhe,~Germany}
\author{Tim Storbeck}
\affiliation{Physikalisches~Institut,~Karlsruhe~Institute~of~Technology,~76131 Karlsruhe,~Germany}
\author{Thomas Gozlinski}
\affiliation{Physikalisches~Institut,~Karlsruhe~Institute~of~Technology,~76131 Karlsruhe,~Germany}
\author{Lukas Gr\"unhaupt}
\affiliation{Physikalisches~Institut,~Karlsruhe~Institute~of~Technology,~76131 Karlsruhe,~Germany}
\author{Ioan M. Pop}
\affiliation{Physikalisches~Institut,~Karlsruhe~Institute~of~Technology,~76131 Karlsruhe,~Germany}
\affiliation{Institute~of~Nanotechnology,~Karlsruhe~Institute~of~Technology,~76344~Eggenstein-Leopoldshafen,~Germany} 
\author{Wulf Wulfhekel}
\affiliation{Physikalisches~Institut,~Karlsruhe~Institute~of~Technology,~76131 Karlsruhe,~Germany}
\affiliation{Institute~of~Nanotechnology,~Karlsruhe~Institute~of~Technology,~76344~Eggenstein-Leopoldshafen,~Germany}

\date{\today}

\begin{abstract}
{\bf Following the emergence of superconducting granular aluminum (grAl) \cite{cohen_superconductivity_1968, deutscher_transition_1973} as a material for high-impedance quantum circuits \cite{grunhaupt_granular_2019,valenti_interplay_2019,hazard_nanowire_2019,zhang_microresonators_2019}, future development hinges on a microscopic understanding of its phase diagram \cite{deutscher_transition_1973,pracht_enhanced_2016, levy-bertrand_electrodynamics_2019}, and whether the superconductor-to-insulator transition (SIT) is driven by disorder \cite{anderson_absence_1958, sacepe_disorder-induced_2008, sacepe_localization_2011, driessen_strongly_2012, mondal_phase_2011} or charging effects \cite{mott_metal-insulator_1968, dynes_tunneling_1984, bachar_kondo-like_2013, bachar_mott_2015, moshe_optical_2019}. Beyond fundamental relevance, these mechanisms govern noise and dissipation in microwave circuits \cite{feigelman_microwave_2018, kuzmin_quantum_2019}. Although the enhancement of the critical temperature, and the SIT in granular superconductors have been studied for more than fifty years \cite{beloborodov_granular_2007}, experimental studies have so far provided incomplete information on the microscopic phenomena. Here we present scanning tunneling microscope measurements of the local electronic structure of superconducting grAl. We confirm an increased superconducting gap in individual grains both near and above the Mott resistivity $\rho_\mathrm{M} \approx \SI{400}{\micro\ohm\,\centi\meter}$ \cite{dynes_metal-insulator_1981, dynes_tunneling_1984}.  Above $\rho_\mathrm{M}$ we find Coulomb charging effects, a first indication for decoupling, and in-gap states on individual grains, which could contribute to flux noise and dielectric loss in quantum devices. We also observe multiple low-energy states outside the gap, which may indicate bosonic excitations of the superconducting order parameter \cite{Sherman2015,Nakamura2019}.}
\end{abstract}

\maketitle

Superconducting granular metals are interesting from a fundamental point of view, because their micro-structure can be seen as a network of artificial atoms coupled by electron tunneling, giving rise to rich physical phenomena \cite{beloborodov_granular_2007}. Beyond this fundamental interest, they are also valuable from a quantum circuit engineering perspective thanks to their large kinetic inductance, low microwave frequency losses \cite{grunhaupt_granular_2019,valenti_interplay_2019,zhang_microresonators_2019} and amenable non-linearity \cite{maleeva_nonlinearity_2018}.
Granular aluminum (grAl) is particularly appealing due to its ease of fabrication and compatibility with standard Al/AlO$_x$/Al Josephson junction technology. Tuning the oxygen partial pressure during aluminum deposition results in a granular structure, film resistivities $\rho$ from \SIrange{10}{e5}{\micro\ohm\,\centi\meter}, and consequently variable inter-grain coupling. As a function of $\rho$, the critical temperature $\Tc$ of grAl increases compared to pure aluminum \cite{cohen_superconductivity_1968}, reaching a maximum in the vicinity of the Mott resistivity $\rho_\mathrm{M} \approx \SI{400}{\micro\ohm\,\centi\meter}$ \cite{dynes_metal-insulator_1981,dynes_tunneling_1984}, after which it decreases, and the system becomes insulating at $\rho \approx \SI{10}{\milli\ohm\,\centi\meter}$.

Although grAl has been successfully employed in fluxonium quantum bits \cite{grunhaupt_granular_2019}, the realization of superconducting quantum processors requires orders of magnitude improvement in coherence. This task is complicated by the fact that several types of imperfections, both microscopic and macroscopic, are concomitantly present, depending on system design, micro-fabrication technology, and materials \cite{krantz_a-quantum_2019}. So far, microscopic defects have been measured rather indirectly, through their interaction with the device itself \cite{Grabovskij2012, wang_measurement_2014}. It is therefore clear that a local probe, such as scanning tunneling microscopy (STM), can offer precious complementary information about the nature of the observed defects, be that localized spins, charge states, or other microscopic systems.

Charging effects due to grain decoupling have been proposed to govern the SIT in grAl, supported by evidence from tunnel junction spectroscopy, magnetoresistance measurements, and muon spin relaxation \cite{bachar_mott_2015,Nimrod2018}. Here, we employ the imaging abilities of a 30~mK STM, capable of high energy-resolution spectroscopy \cite{Balashov18}, to resolve localized charge states and to test the presence of both in-gap states and low energy excitations outside the gap.


\begin{figure}[htb]
  \centering
  \includegraphics[width=0.9\columnwidth]{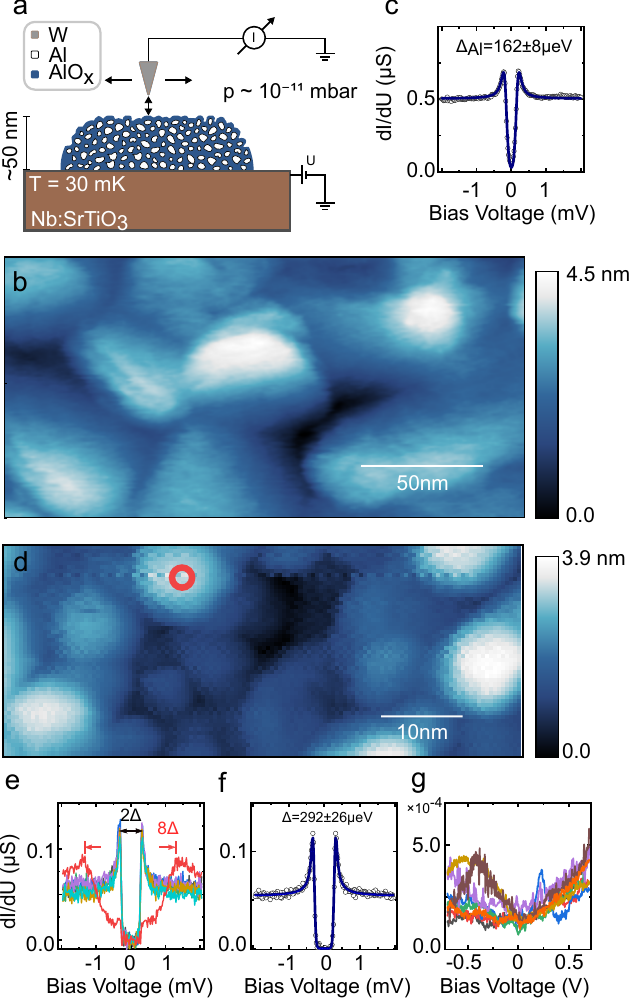}\\
  \caption{{\bf Topographic and electronic structure of pure Al, and of oxygen poor grAl films with $\rho \approx \SI{300}{\micro\ohm\centi\meter}$.} 
  (a) Schematic drawing of the experimental configuration, consisting of the substrate, the thin grAl film, and the STM tip. (b) Topographic image of a pure Al film ($U$=-1\ V, $I$=100\ pA). (c) Differential conductivity measured on the pure Al film showing a superconducting gap $\Delta$ close to that of bulk Al ($U$=-2\ mV, $I$=1\ nA, ${\Delta}U_{\mathrm{rms}}= \SI{14}{\micro\volt}$).
  (d) Topographic image of a grAl film ($U$=60\ mV, $I$=240\ pA). 
  (e) Differential conductance $dI/dU$ recorded on different grains. All measured grains show the same gap $\Delta \approx \SI{290}{\micro\electronvolt}$ ($U$=2\ mV, $I$=107\ pA, ${\Delta}U_{\mathrm{rms}}=\SI{15}{\micro\volt} $). The red curve was recorded on the grain marked with a red circle in (d), in which besides the superconducting gap, a secondary gap of $\approx 8 \Delta$ was observed. (f) Averaged spectrum (open circles) of (e), excluding the red curve, and BCS fit (blue line) including temperature and modulation broadening. (g) Large bias range $dI/dU$ measured on different grains, showing essentially a metallic behaviour ($T=$1\ K, $U$=700\ mV, $I$= 156\ pA, ${\Delta}U_{\mathrm{rms}}$= 3\ mV).}
  \label{Fig1}
\end{figure}

The measurement geometry of the experiment is sketched in Fig.~\ref{Fig1}a. A conducting Nb doped SrTiO$_3$ (Nb:STO) single crystal substrate is used to apply the sample bias $U$ to the film, and the STM tip is scanned over its surface. For topographic images, the bias voltage is selected to be much above the superconducting gap and the STM is operated in the constant current mode, in which the tip is regulated in the z-direction by a feed-back loop such that the tunneling current $I$ at the chosen bias voltage is constant. To obtain information on the superconducting gap, the feed-back loop is opened, the position of the tip is fixed and the bias voltage is ramped, while the differential conductivity is measured.

As a control experiment, we start by measuring a \SI{30}{\nano\meter} thick pure Al film. 
Figure \ref{Fig1}b shows a typical STM image of the film, which consists of crystallites of about 50~nm lateral size. The large size of the crystallites is due to the high diffusion mobility of Al, according to its relatively low melting point.
Figure \ref{Fig1}c shows the superconducting gap and the coherence peaks in the Al spectrum. The superconducting gap $\Delta_\mathrm{Al}$ of about \SI{160}{\micro\electronvolt}, obtained by a fit to the BCS density of states (DOS), is close to the known values for thin-film Al \cite{Douglas64}. In all respects, the pure Al sample behaves as expected.

\begin{figure*}[thb]
  \centering
  \includegraphics[width=2\columnwidth]{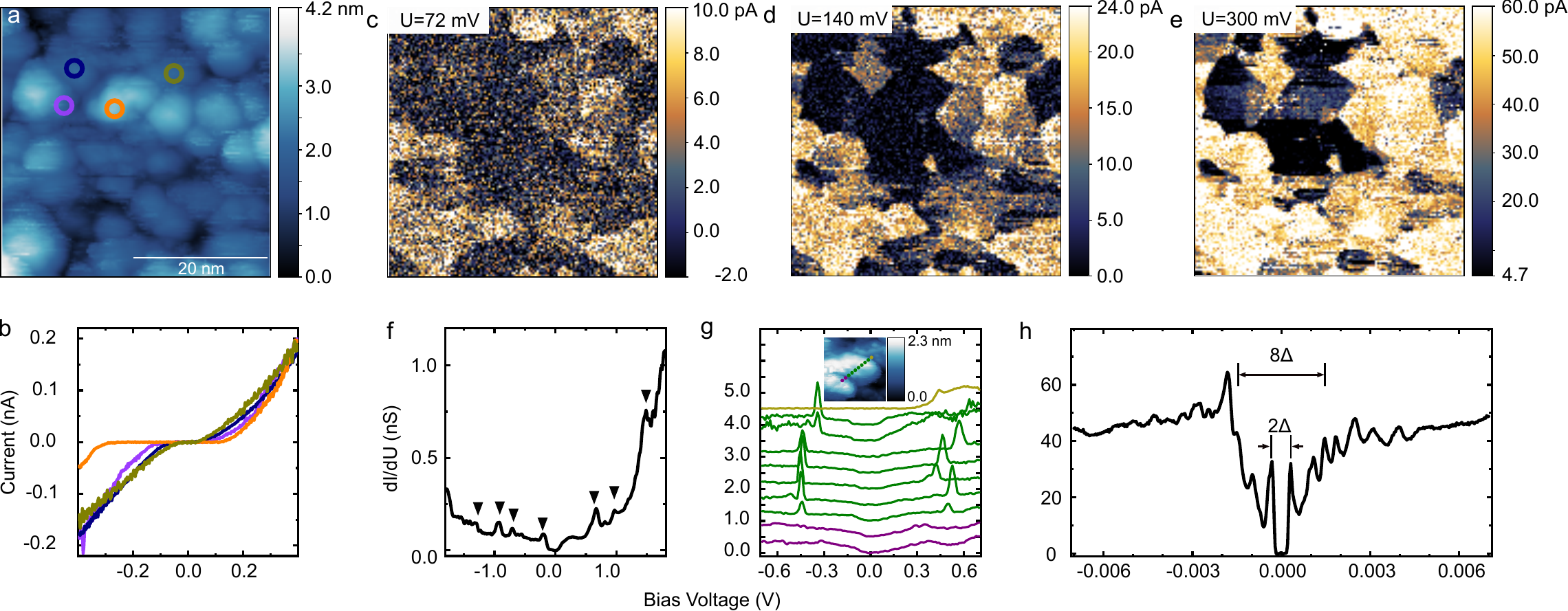}\\
  \caption{{\bf Topographic and electronic structure of oxygen rich grAl films with $\rho\approx\SI{2e3}{\micro\ohm\centi\meter}$.} (a) Topographic image of the grAl film ($U$=400\ mV, $I$=180\ pA). (b) Averaged $I(U)$ spectra of grains marked by the four colored circles in (a) with same feedback parameters. (c-e) Maps of the local tunneling current $I$ for the same area as in (a) at positive bias voltages, as indicated by the labels. Abrupt changes of the contrast in several grains at the same voltage indicate electrical connection between the grains (tip stabilized at: $U$=400\ mV, $I$=180\ pA).  (f) Typical $dI/dU$ spectrum of an individual grain at energies much higher than the superconducting gap showing peaks -- marked by triangles -- indicative for Coulomb blockade effects ($U$=1.8\ V, $I$= 550\ pA, $U_\mathrm{rms}$=21\ mV). (g) Series of $dI/dU$ spectra recorded on a line crossing three grains (c.f. topographic inset -- 20$\times$20 nm$^2$) indicated by purple, green and yellow (neighboring curves are shifted by \SI{0.5}{\nano\siemens}). The position of the charging peaks varies slightly within a grain and changes abruptly from one grain to another ($U$=700\ mV, $I$=200\ pA, ${\Delta}U_{\mathrm{rms}}$=7\ mV). (h) Typical high resolution $dI/dU$ spectrum of a grain, which, besides the superconducting gap of grAl centered around the Fermi energy, shows sharp peaks (excitations) within a secondary gap of $\approx 8\Delta$ ($U$=7\ mV, $I$=270\ pA, ${\Delta}U_{\mathrm{rms}}=\SI{14}{\micro\volt}$).} \label{Fig2}
\end{figure*}

In contrast, Fig.~\ref{Fig1}d shows a typical topographic image of the sample surface of a grAl film with $\rho \approx \SI{300}{\micro\ohm\centi\meter}$, i.e., close to the Mott resistivity and the maximum $T_C=2.1$~K of the superconducting dome~\cite{levy-bertrand_electrodynamics_2019}. Superconducting grAl can be viewed as a connected 3D network of Al grains, separated by amorphous AlO$_x$ oxide (for film deposition details see Methods). As evident from Fig.~\ref{Fig1}d, the Al grains in grAl are much smaller compared to the pure Al film, as expected \cite{deutscher_transition_1973}.
Note that the STM tip, etched from a W wire, cannot be made to arbitrary sharpness, and has a typical radius of curvature in the tens of nm range. Since the topographic image reflects a convolution of the grains and the tip shape, the grains may appear slightly larger in STM images. The observed distribution of grain sizes \SIrange{5}{10}{\nano\meter}, however,  agrees with the reported values and spread in the literature for similar film growth conditions \cite{deutscher_transition_1973}.

Figure \ref{Fig1}e shows a selection of differential conductance $dI/dU$ spectra recorded on different grains. They show a practically identical superconducting gap $\Delta$, yielding a value for the grAl superconducting gap $\Delta=292\pm26$~\SI{}{\micro\electronvolt} obtained with a BCS fit (see Fig.~\ref{Fig1}f). Note that this value is significantly larger than that of pure Al (cf. Fig.~\ref{Fig1}c). For similar grAl films, radio-frequency measurements \cite{valenti_interplay_2019} revealed a bulk value for $\Delta$ of approximately $\SI{344}{\micro\electronvolt}$, extracted from the measurement of the critical temperature $T_C$, while direct THz spectroscopy \cite{levy-bertrand_electrodynamics_2019} revealed a value of 
$336\pm8$~\SI{}{\micro\electronvolt}. The surprisingly small difference between these values, obtained using very different methodologies, indicates that grains residing at the surface have a superconducting gap comparable to the bulk of the grAl sample. Additionally, from the fact that all measured grains show the same $\Delta$, we conclude that either the gap enhancement mechanism is homogeneous and robust to variations in grain size, or the grains are coupled strongly enough to even out $\Delta$ variations. 

On a larger energy scale (see Fig.~\ref{Fig1}g) the grains show a metallic behaviour, i.e. a DOS that has no conventional band gap. However, the $dI/dU$ spectra are quantitatively different from grain to grain. The fact that we do not observe a common DOS over several grains indicates that the electron wave functions are not entirely delocalized and electrons are partly reflected at the boundaries, which is expected considering the grains are oriented randomly and are separated by amorphous AlO$_x$ barriers. 
\begin{figure}[htb]
  \centering
  \includegraphics[width=\columnwidth]{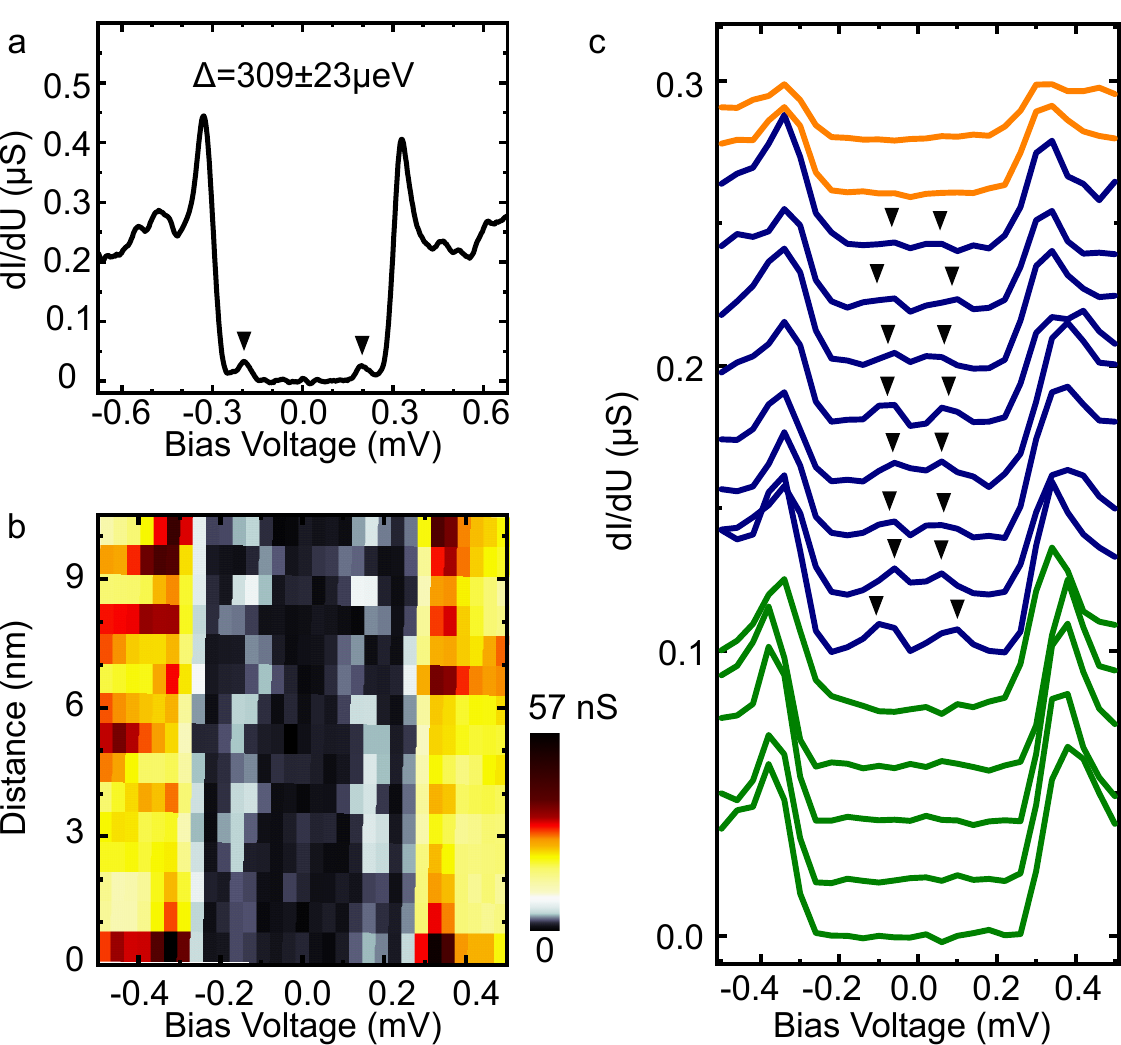}\\
  \caption{{\bf In-gap states in oxygen rich grAl films with $\rho \approx \SI{2e3}{\micro\ohm\centi\meter}$.} (a) High-resolution $dI/dU$ spectrum showing two in-gap states, highlighted by triangular markers, at $\pm\SI{200}{\micro\volt}$ ($U$=0.7\ mV, $I$=10\ pA, energy resolution \SI{20}{\micro\volt}).  (b) Colour coded differential conductivity as a function of lateral tip position, recorded on a straight line, across one grain. The in-gap states shift symmetrically in energy with respect to the origin as a function of tip position ($U$=4\ mV, $I$=100\ pA, ${\Delta}U_{\mathrm{rms}}=\SI{20}{\micro\volt}$). (c) A series of $dI/dU$ spectra recorded in a straight 18~nm line crossing three grains indicated by green, blue and orange. The two outer grains do not show in-gap states, while the inner grain displays in-gap states at $\approx\pm \SI{100}{\micro\volt}$ ($U$=4\ mV, $I$=100\ pA, ${\Delta}U_{\mathrm{rms}} =\SI{25}{\micro\volt}$). }\label{Fig3}
\end{figure}
Note that occasionally, individual grains show a deviating behaviour, as indicated by the red circle in Fig.~\ref{Fig1}d and the corresponding red $dI/dU$ spectrum in Fig.~\ref{Fig1}e. 
The low energy spectrum of the grain shows reduced coherence peaks, and a secondary gap of approximately $8\Delta$. These are the first indications of what will be observed in grAl films with a higher oxygen concentration.

In Fig.~\ref{Fig2}, we summarize the behaviour of grains in oxygen rich films ($\rho \approx \SI{2e3}{\micro\ohm\centi\meter}$). In topography (see Fig.~\ref{Fig2}a) we observe grains of similar size compared to the oxygen poor films.  We also notice sudden jumps in the z-position when scanning some of the grains. These indicate charging effects, as also observed on a few, more insulated grains in Fig.~\ref{Fig1}d. From the large range $I(U)$ measurements shown in Fig.~\ref{Fig2}b, taken on the individual grains highlighted in Fig.~\ref{Fig2}a, it becomes obvious that the grains do not entirely behave metallic, but only show large currents above (or below) a certain threshold voltage. The $I(U)$ curves also vary from grain to grain.
In Fig.~\ref{Fig2}, panels c-e, we plot measured maps of the tunnel current at several bias voltages (\SI{72}{mV}, \SI{140}{mV}, and \SI{300}{mV}). Multiple neighbouring grains show a sudden increase in current, indicating that they form stronger electrically coupled clusters. Note that we measure the same behaviour for both polarities of the bias voltage (see Supplementary Information). 
Moreover, charging effects are also apparent in the differential conductance in Fig.~\ref{Fig2}f. Sudden changes in current correspond to peaks in $dI/dU$, marked by black triangles. Such series of peaks are common in Coulomb blockade \cite{10.3389/fphy.2013.00013}.

We also observe charging localized on a single grain. In Fig.~\ref{Fig2}g we show a series of $dI/dU$ curves recorded on a line extending over three grains. Measurements on each grain are indicated by purple, green and yellow spectra. The features of the spectra are qualitatively similar when the STM tip probes the same grain, however, in between the grains, the spectra change discontinuously.

Since STM is only sensitive to the surface, one can speculate that grains inside the film are coupled to more neighbors, exhibiting less charging. However, for films with thickness comparable to the grain size, charging effects could be dominant. Indeed, recent measurements of a grAl transmon quantum bit, employing a \SI{10}{nm}~thick film, also suggest that grains couple in clusters \cite{winkel_grAl_transmon_2019}, comparable to the ones visible in Fig.~\ref{Fig2}c-e.

From the measured low energy spectrum in Fig.~\ref{Fig2}h, we confirm that the grains are superconducting, with a similarly enhanced superconducting gap $\Delta$ (see Supplementary Information for spectra at higher temperatures). The current density used for the measurement of the low-energy tunneling spectra is three orders of magnitude below the bulk critical current density of similar films \cite{Friedrich_2019}, so we expect the measurement to be nondisruptive. Nevertheless, for oxygen rich films we observe a secondary gap $\approx 8\Delta$ (cf. Fig.~\ref{Fig2}h), and a series of peaks outside $\Delta$ but inside the secondary gap. We can only speculate on the origin of this presumably many body effect. The energy and sharpness of the peaks are distinctly different from the charging effects shown in Fig.~\ref{Fig2}g, and resemble a repetition of the coherence peaks. This might indicate inelastic tunneling, i.e. the energy of the tunneling electron is shared between an excited boson, of defined energy, and the scattered electron ending up in the coherence peak. Possible bosonic excitations are Higgs modes in the granular superconductor \cite{Sherman2015,Nakamura2019} or plasmons \cite{levy-bertrand_electrodynamics_2019}. As the spacing of the peaks is not regular, and they are not symmetric with respect to zero bias, this hypothesis needs to be further explored, both theoretically and experimentally.

We now focus on the superconducting gap in oxygen rich samples. Figure~\ref{Fig3}a shows a high resolution spectrum of the superconducting gap with $\Delta=309\pm23$~\SI{}{\micro\electronvolt} obtained from a BCS fit. 
The gap $\Delta$ measured on different grains agrees within the error bar, and its value is slightly higher compared to the oxygen poor samples (cf. Fig.~\ref{Fig1}f). However, about half of the grains show in-gap states, as indicated by triangles in the $dI/dU$ spectrum of Fig.~\ref{Fig3}a. These states are of rather low intensity, but their energetic position rules out Andreev reflections of the tip electrons as their origin. Instead, the observation of the Kondo effect in grAl films above $T_C$ \cite{Moshe2018} 
suggests that unpaired electrons or magnetic moments give rise to Yu-Shiba-Rusinov (YSR) in-gap states. The energy of YSR states is a continuous function of the exchange coupling to the superconductor \cite{Franke2018}, and of the charging energy of the grain or cluster of grains.

Similarly to Ref.~\cite{bachar_kondo-like_2013}, we consider two scenarios for the origin of the unpaired spins. First, the finite size of the grains and the fact that some are only weakly coupled to the rest of the grAl film might result in an odd number of charges on individual grains in equilibrium, as suggested by Deutscher et al. \cite{Moshe2018}.
The interaction of these unpaired electrons with the superconducting condensate may induce YSR states. Second, the non-stoichiometric composition of the AlO$_x$ in-between the pure Al grains may lead to unpaired electrons trapped in the insulating barriers. The interaction of these very localized states with the superconductor could be responsible for the measured YSR states. 

In order to discriminate between the above scenarios, we measured the spacial dependence of the YSR states on a line crossing a single grain. While the superconducting gap $\Delta$ remains constant, as shown in Fig.~\ref{Fig3}b, the in-gap states vary in energy as a function of the STM tip position. In the scenario of electrically decoupled grains with an odd number of electrons, the electric field of the STM tip would change the chemical potential of the grain, the coherence peaks would shift with tip position, but the energy of the YSR states with respect to the gap edge would be fixed. This is in contrast with our observations. If the unpaired spins reside in the oxide, the electric field of the tip can change their energy, and consequently the exchange coupling to the grain, as in the study of Farinacci et al. \cite{Franke2018}. In this case the YSR state shifts with tip position, while the coherence peaks remain unaffected. This scenario agrees with our measurements.
Moreover, Fig.~\ref{Fig3}c illustrates the strong localization of the YSR states on particular grains. Only the middle grain (blue spectra) shows in-gap states at $\pm\SI{100}{\micro\electronvolt}$ (triangle markers).

In summary, using STM spectroscopy we have observed an enhanced and isotropic superconducting gap for both oxygen poor and rich grAl samples. In the oxygen rich samples the grains start to decouple, charging effects set in, eventually leading to the SIT. Hand in hand with this decoupling, we observed a secondary gap and excitations with an energy comparable to the superconducting gap, as well as YSR states in some of the grains, possibly related to unpaired electrons in the oxide barriers. For the YSR states to exist, the localized spins in the oxide need to exchange couple with the superconducting grain. If the oxide is sufficiently thin, the states in the oxide delocalize and we do not observe in-gap states. This interpretation is consistent with the fact that the YSR states appear only when the grains start to decouple, for resistivities larger than the Mott resistivity $\rho_\mathrm{M}$. 

The existence of in-gap YSR states can be detrimental for grAl quantum devices in several ways: they can introduce additional dissipation in the microwave domain, or even alter the spectrum of the devices, if they couple sufficiently strong \cite{Grabovskij2012}. The presence of spins may also contribute to the flux noise observed in grAl devices \cite{grunhaupt_granular_2019}, and more generally in pure Al devices, where these spins can form in the natural oxide at the surface and in the Josephson junction barrier. Further work should focus on measuring the density of YSR states versus resistivity in grAl, and also in native Al oxide, with the goal of reducing their impact on coherence in quantum devices.


\bibliographystyle{nature}
\bibliography{AlO.bib}


\vspace{2mm}

\noindent
{\bf Methods}
The samples consists of \SI{50}{\nano\meter} thick grAl films deposited by electron beam evaporation of pure Al at a rate of \SI{1}{\nano\meter\per\second} in an oxygen atmosphere, at a pressure in the \SI{e-5}{\milli\bar} range (c.f. Refs.~\cite{deutscher_transition_1973, grunhaupt_granular_2019, levy-bertrand_electrodynamics_2019}). During the deposition the substrate is at room temperature. The resistivity of the films used in this work is in the \SI{300}{\micro\ohm\centi\meter} to \SI{2000}{\micro\ohm\centi\meter} range, corresponding to a superconducting critical temperature near \SI{2}{\kelvin} (c.f. SI). For these types of films, the SIT is observed for a room temperature resistivity $\rho \approx \SI{10}{\milli\ohm\centi\meter}$. We used Nb doped (0.7~weight~\%) StTiO$_3$ (Nb:STO) single crystals as substrates, in order to start off with a conducting, flat, and non-reactive surface. The substrate is expected to become superconducting at temperatures below 200~mK \cite{Pfeiffer_1969}. Since the film thickness is much larger than the grain size of \SIrange{3}{5}{\nano\meter} \cite{deutscher_transition_1973}, the films show a bulk-like behaviour, i.e. most of the grains reside inside the film and determine the transport properties. It is therefore not surprising that the measured properties of grAl films are consistent between various substrates, such as glass, silicon or sapphire \cite{deutscher_transition_1973, levy-bertrand_electrodynamics_2019}. Immediately after deposition, in order to avoid surface oxidation of the grAl films, the samples were transferred to the ultra high vacuum of the STM using a vacuum suitcase, at a pressure in the \SI{e-9}{\milli\bar} range. STM measurements were carried out at a base temperature of 30 mK. If other temperatures were used, it is indicated in the figure captions. 

\vspace{2mm}

\noindent
{\bf Acknowledgements}
W.W. acknowledges funding by the Deutsche Forschungsgemeinschaft (DFG) under the grant WU 349/12-1 and INST 121384/30-1 FUGG. I.M.P and L.G. acknowledge the Alexander von Humboldt Foundation in the framework of a Sofja Kovalevskaja award endowed by the German Federal Ministry of Education and Research. F.Y. acknowledges funding from the Alexander von Humboldt foundation.  

\noindent
{\bf Author contribution}
F.Y., I.M.P. and W.W. conceived the experiments. F.Y., T.S. and T.G. carried out the STM experiments and analyzed the data. I.M.P. and L.G. grew the samples.  F.Y., W.W., L.G., and I.M.P. wrote the draft. All authors contributed in discussions and finalizing the manuscript.


\end{document}



\title{Supplementary Information for\\
``Microscopic charging and in-gap states \\ in superconducting granular aluminum''}


\author{Fang Yang}
\affiliation{Institute~for~Nanoelectronic~Devices~and~Quantum~Computing,~Fudan~University,~200438~Shanghai,~China}
\affiliation{Physikalisches~Institut,~Karlsruhe~Institute~of~Technology,~76131 Karlsruhe,~Germany}
\author{Tim Storbeck}
\affiliation{Physikalisches~Institut,~Karlsruhe~Institute~of~Technology,~76131 Karlsruhe,~Germany}
\author{Thomas Gozlinski}
\affiliation{Physikalisches~Institut,~Karlsruhe~Institute~of~Technology,~76131 Karlsruhe,~Germany}
\author{Lukas Gr\"unhaupt}
\affiliation{Physikalisches~Institut,~Karlsruhe~Institute~of~Technology,~76131 Karlsruhe,~Germany}
\author{Ioan M. Pop}
\affiliation{Physikalisches~Institut,~Karlsruhe~Institute~of~Technology,~76131 Karlsruhe,~Germany}
\affiliation{Institute~of~Nanotechnology,~Karlsruhe~Institute~of~Technology,~76344~Eggenstein-Leopoldshafen,~Germany} 
\author{Wulf Wulfhekel}
\affiliation{Physikalisches~Institut,~Karlsruhe~Institute~of~Technology,~76131 Karlsruhe,~Germany}
\affiliation{Institute~of~Nanotechnology,~Karlsruhe~Institute~of~Technology,~76344~Eggenstein-Leopoldshafen,~Germany}

\date{\today}

\begin{abstract}
{\bf The supplementary information consists of the following: the current maps at negative bias voltage, tunneling spectra at different tunneling conductance, and high energy-resolution spectra at higher temperatures.}
\end{abstract}

\maketitle

\newpage

\section{\bf{Current maps at negative bias voltage}}

\begin{figure*}[h]
  \centering
  \includegraphics[width=\columnwidth]{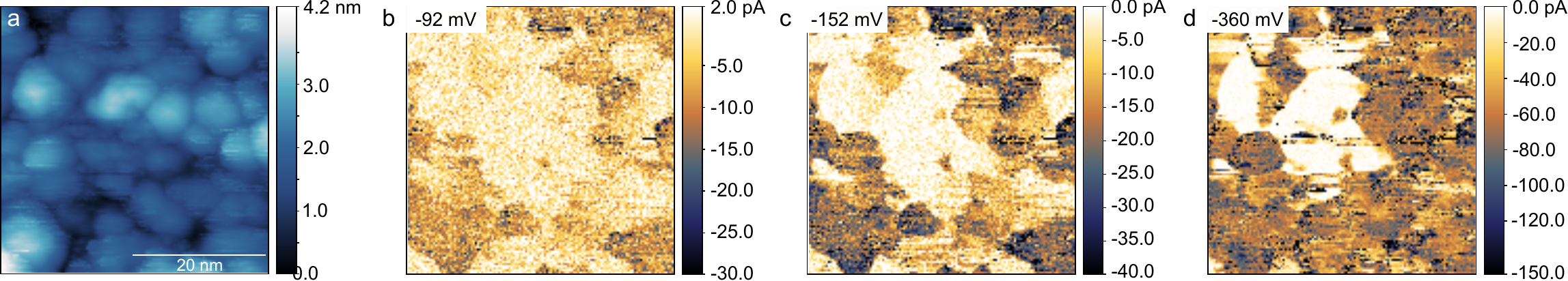}\\
  \caption{{\bf Topographic and negative current images.} 
  (a) Topographic image of the granular film, the same as Fig. 2a of the main text. (b-c) Maps of the local tunneling current $I$ of this area at negative bias voltages, as indicated by the labels.}\label{FigS1}
\end{figure*}

In Fig. 2 of the main text, we show the current maps at positive bias voltages together with topography in the oxygen-rich sample. In Fig. \ref{FigS1}b-c we show the corresponding current maps for negative bias voltages, as indicated. Similar to the observation at positive bias, the grains turn dark in large groups with increasing negative bias voltages. The grain clustering is similar for both polarities.

\newpage

\section{\bf{Measurements at different tunneling conductance}}

\begin{figure*}[h]
  \centering
  \includegraphics[width=0.5\columnwidth]{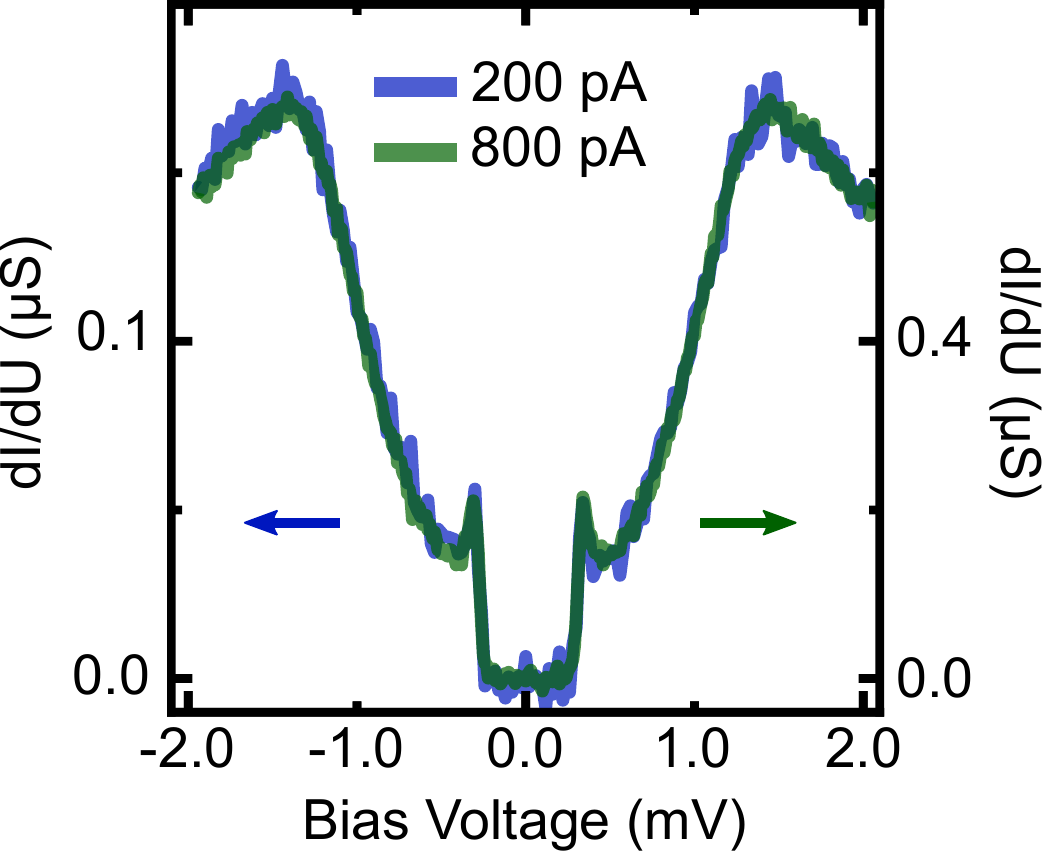}\\
  \caption{{\bf Measurements $dI/dU$ with different tunneling conductance, on a relatively insulated grain of the oxygen-poor sample.} On the grain marked by the red circle in Fig.1d, $dI/dU$ with different tunneling conductance were measured. The feed-back condition for the blue curve was set to be $U=2$\ mV, $I=200$\ pA, and for the green curve  $U=2$\ mV, $I=800$\ pA. For both, $U_{\mathrm{rms}} = \SI{21}{\micro\volt}$.} \label{FigS2}
\end{figure*}

In STM experiments the signal current flows through the tip-sample junction. In our measurements, the size and position of the reported features do not depend on the tunneling conductance, as illustrated in Fig. \ref{FigS2}. The two $dI/dU$ curves in Fig. \ref{FigS2} were measured on the grain marked by a red circle in Fig. 1d of the main text. The shape of the curves is identical, despite the factor four change in the tunneling conductance, i.e. only sample properties enter the $dI/dU$ curves, and there are no detectable voltage drops inside the granular film. We conclude that the tunneling resistance of the tip-sample junction dominates over other resistors in series in the electronic circuit.

\newpage
\section{\bf{High energy-resolution spectra at higher temperatures}}

\begin{figure*}[h]
  \centering
  \includegraphics[width=0.5\columnwidth]{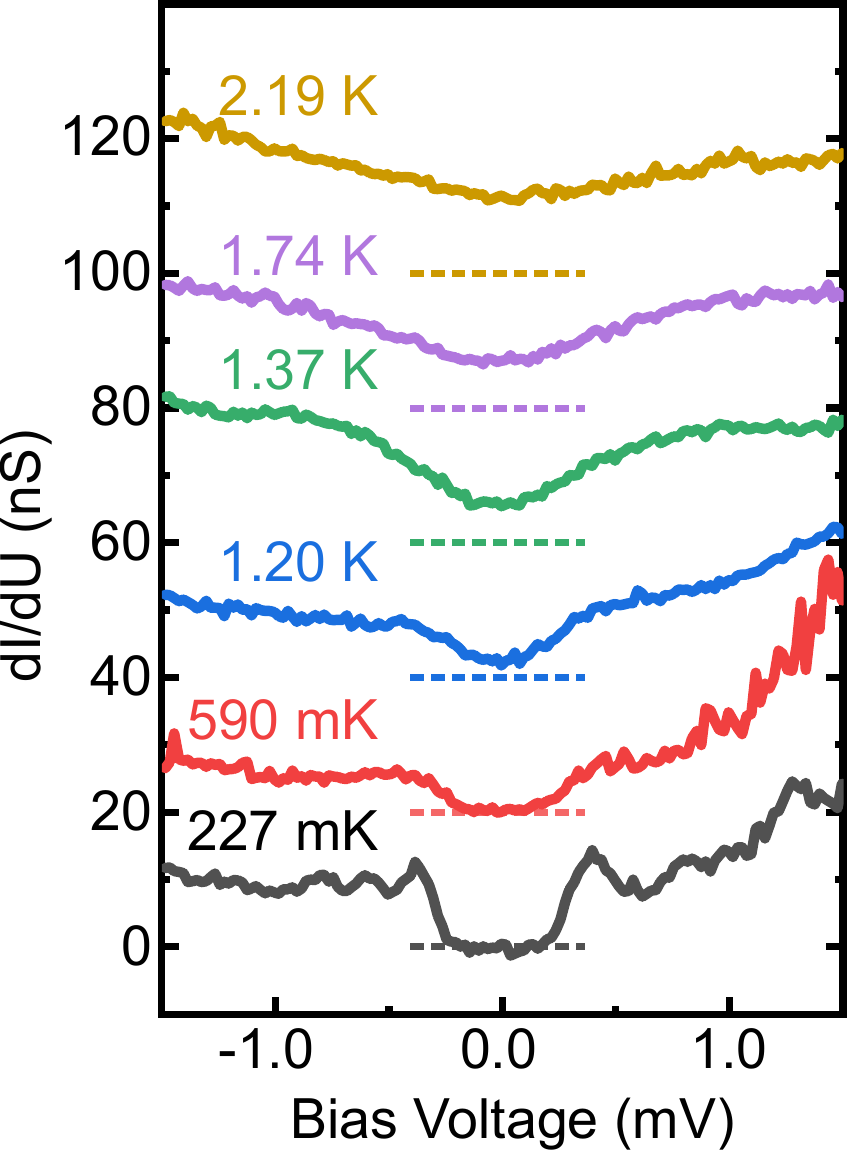}\\
  \caption{{\bf Measured $dI/dU$ for the oxygen-rich sample at different temperatures.} The tunneling conductance was recorded when the sample was warming up. The offset of each spectrum is indicated by dashed lines in corresponding colors. The feed-back condition was set to be $U=2$\ mV, $I=30$\ pA, $U_{\mathrm{rms}} = \SI{70}{\micro\volt}$.} \label{FigS4}
\end{figure*}

We measured the superconducting gap of the oxygen-rich sample (cf. main text) at different temperatures. As shown in Fig. \ref{FigS4}, the superconducting gap smears out with increasing temperature and eventually vanishes. From the limited number of measurements at various temperatures, we can only state that the superconducting transition occurs between \SI{1.74}{\kelvin} and \SI{2.19}{\kelvin}

\newpage

%

\bibliographystyle{nature}
\bibliography{AlO.bib}